\documentclass[prl,showpacs,twocolumn,amsmath,amssymb]{revtex4}

\usepackage{mathrsfs}%hanamoji
\usepackage[dvipdfmx]{graphicx}% Include figure files
\usepackage{dcolumn}% Align table columns on decimal point
\usepackage{bm}% bold math
\usepackage{color}
\usepackage{comment}
\usepackage{enumitem}
\usepackage{ulem}

\newcommand{\tl}{\tilde}
\newcommand{\ra}{\rangle}
\newcommand{\la}{\langle}
\newcommand{\tb}{\textbf}

\newcommand{\hxi}{\hat{\xi}}

\newcommand{\hz}{\hat{z}}

\newcommand{\mrP}{\mathrm{P}}

\newcommand{\mrss}{\mathrm{ss}}

\newcommand{\htt}{\hat{t}}

\newcommand{\hnu}{\hat{\nu}}
\newcommand{\hN}{\hat{N}}
\newcommand{\Lpath}{\mathcal{L}_{\mathrm{path}}}
\begin{document}

\title{Non-universal power law distribution of intensities of the self-excited Hawkes process: a field-theoretical approach}

\author{Kiyoshi Kanazawa$^{1}$ and Didier Sornette$^{2-4}$}

\affiliation{
	$^1$ Faculty of Engineering, Information and Systems, The University of Tsukuba, Tennodai, Tsukuba, Ibaraki 305-8573, Japan\\
	$^2$ ETH Zurich, Department of Management, Technology and Economics, Zurich, Switzerland\\
	$^3$ Tokyo Tech World Research Hub Initiative, Institute of Innovative Research, Tokyo Institute of Technology, Tokyo, Japan\\
	$^4$ Institute of Risk Analysis, Prediction and Management, Academy for Advanced Interdisciplinary Studies, Southern University of Science and Technology, Shenzhen, China
}
\date{\today}

\begin{abstract}
	The Hawkes self-excited point process provides an efficient representation of the bursty intermittent dynamics of many
	physical, biological, geological and economic systems. By expressing the probability for the next event per unit time
	(called ``intensity''), say of an earthquake, 
	as a sum over all past events of (possibly) long-memory kernels, the Hawkes model is non-Markovian. 
	By mapping the Hawkes model onto stochastic partial differential equations that are Markovian, we develop a field theoretical approach in terms of probability density functionals.
	Solving the steady-state equations, we predict a power law scaling of the probability density function (PDF) of the intensities close to the critical point $n=1$ of the Hawkes process,
	with a non-universal exponent, function of the background intensity $\nu_0$ of the Hawkes intensity, the average time scale
	of the memory kernel and the branching ratio $n$. Our theoretical predictions are confirmed by numerical simulations.
	\end{abstract}

\maketitle
	Most out-of-equilibrium dynamical processes in physical, natural and social systems 
	are characterised by the presence of extended and often long memory. 
	A prominent class of such non-Markovian dynamics includes epidemic spreading processes, which have broad applications 
	from photoconductivity in amorphous semiconductors and organic compounds \cite{ScherMontroll75},
	rainfall and runoff in catchments \cite{Scheretal2002}, earthquake interactions \cite{HelmsSor02},
	epidemiology \cite{Feng-epidemic2019}, brain memory \cite{Andersonbrain01},
	credit rating \cite{DAmicoetal19} and default cascades in finance \cite{Errais-Giesecke2010}, 
	financial volatility dynamics \cite{Chakraetal11,Jiangealmultifract19}, transaction intervals in foreign exchange markets \cite{Taka2},  social dynamics
	of book sales  \cite{SorDeschatres04} and YouTube videos views \cite{CraneSor08}, to cite a few \cite{DSendoreview05}.
	
	The self-excited conditional Poisson process introduced by Hawkes \cite{Hawkes1,Hawkes2,Hawkes3} 
	is the simplest point process modelling epidemic dynamics, in which the whole past 
	history influences future activity. It captures the ubiquitous phenomenon of time (and space)
	intermittency and clustering due to endogenous interactions.  The Hawkes process is enjoying
	an explosion of interest in many complex systems, including
	in physics, biology, geology and seismology and in financial and economic markets. For instance,
	the Hawkes model remains the standard reference in statistical seismology 
	\cite{Ogata1988,Ogata1999,HelmsSor02,Shyametal2019} and  is now used
	to model a variety of phenomena in finance, from microstructure dynamics to default risks \cite{FiliSor12,HawkesRev18}.
	The Hawkes process is also fashionable to model social dynamics \cite{Zhao2015}.
	
	A key ingredient of the Hawkes model is the memory kernel $h(t)$, which quantifies
	how much a past event influences the triggering of a future event. When $h(t)$ is a pure
	exponential function, the Hawkes model can be represented as a Markovian process by adding an auxiliary variable.
	But most systems exhibit longer memories, with $h(t)$ containing multiple time scales and often 
	describing power law decaying impacts, which makes the Hawkes model non-Markovian in general.
	Here, we present a general field master equation, which represents the self-excited Hawkes process
	as being equivalent to a stochastic Markovian partial differential equation. 
	This novel representation allows us to use the mathematical apparatus to solve master equations
	and derive a new result on the distribution of  activity rates, which is found to take the form of a non-universal power law.
	
	The Hawkes process is defined via its intensity $\hnu$, which is the frequency of events per unit time.
	An event can be a burst of electrons in a semiconductor, a rainfall, an earthquake, an epidemic, 
	an epileptic seizure, a firm's bankruptcy or credit default, a financial volatility burst, the sale of a 
	commercial product, viewing a video or a movie, a social action, and so on.
	\begin{figure*}
		\centering
		\includegraphics[width=150mm]{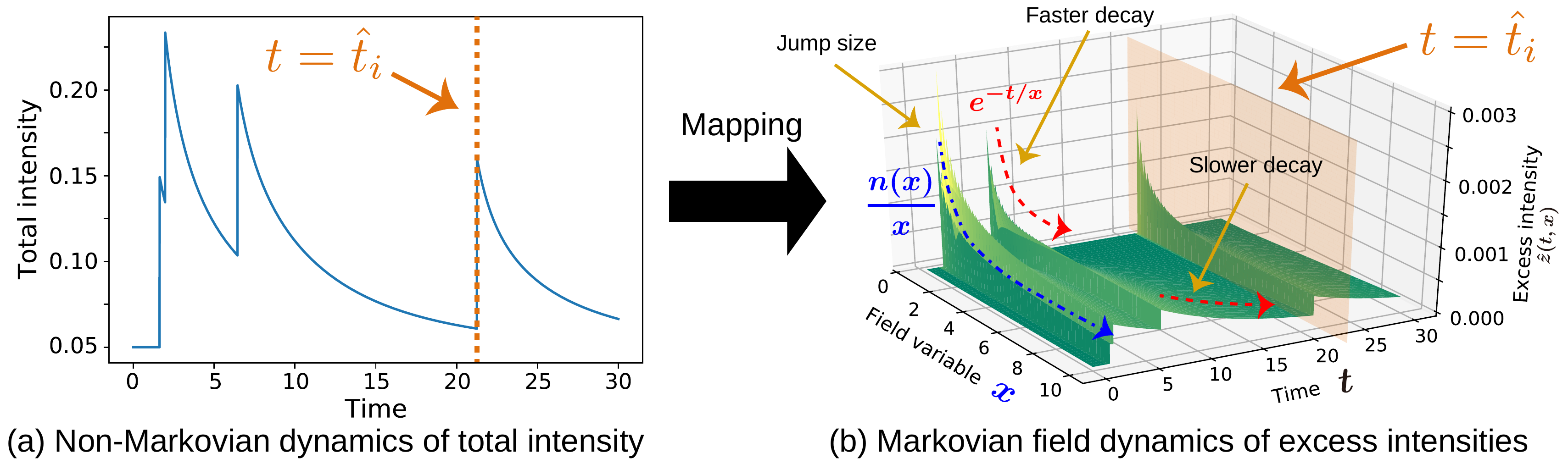}
		\caption{
			Mapping from (a) a non-Markovian description of the total intensity $\hnu(t)$, obeying the original Hawkes process~\eqref{def:Hawkes_general}, to 
			(b) a Markovian description of the excess intensity $\{\hz(t,x)\}_x$ on the auxiliary field variable $x \in \bm{R}_+$, 
			which obeys a SPDE~\eqref{eq:SPDE_generalHawkes}. 
			Notice the shocks~\eqref{hwtrgfq} impact all excess intensity $\{\hz(t,x)\}_{x}$ for all different variables $x$ simultaneously 
			(orange plane at $t=\hat{t}_i$).
			One can observe the dependence $n(x)/x$ of the jump size given in Eq.~\eqref{eq:SPDE_generalHawkes} along the field variable $x$ (blue chain arrow).
			Along the time axis, the exponential decay $\sim e^{-t/x}$ is shown to be faster (slower) for smaller (larger) $x$'s (red broken arrow).
			The sample trajectory was generated with $n(x)=c_{\rm ini}(x-x_{\rm ini})+n_{\rm ini}$ 
			if $x \in [x_{\rm ini}, x_{\rm fin}]$ ($n(x)=0$ otherwise), 
			with $\nu_0=0.05$, $x_{\rm ini}=0.5$, $x_{\rm fin}=10.0$, $c_{\rm ini}=0.002/\Delta x$, 
			$n_{\rm ini}=0.001/\Delta x$, $\Delta x=(x_{\rm fin}-x_{\rm ini})/200$, and $\Delta t=0.006$.
		}
		\label{fig:Mapping}
	\end{figure*}
	Such an event (or shock) occurs during $[t,t+dt)$ with the probability of $\hnu dt$, with
	\begin{align}
		\hnu(t) = \nu_0 + n \sum_{i=1}^{\hN(t)} h(t-\htt_i), \label{def:Hawkes_general}
	\end{align}
	where $\nu_0$ is the background intensity, $\htt_i$ represents the time series of events, 
	$n$ is a positive number called branching ratio, $h(t)$ is a normalized nonnegative function satisfying $\int_0^\infty h(t)dt=1$,
	and $\hN(t)$ is 
	the number of events during the interval $[0,t)$ (called ``counting process''), as shown in Fig.~\ref{fig:Mapping}a for a schematic. 
	By convention, we denote stochastic variables with a hat symbol, such as $\hat{A}$, to distinguish them from
	the non-stochastic real numbers $A$, corresponding for instance to a specific realisation of the random variable. 
	The memory kernel $h(t)$ represents the usually non-Markovian influence of a given event.
	The branching ratio $n$ is the average number of events of first generation (``daughters'') triggered by a given event \cite{DalayVere03,HelmsSor02}
	and is also the fraction of events that are endogenous, i.e., that have been triggered by previous events \cite{HelmsSor03}.
	The Hawkes process has three different regimes: (i) $n<1$: subcritical; 
	(ii) $n=1$: critical and (iii) $n>1$: super-critical or explosive (with a finite probability). 
	The Hawkes process is a model for out-of-equilibrium systems without detailed balance~\cite{GardinerB} and does not satisfy the fluctuation-dissipation relation~\cite{KuboB}.
	
	Let us decompose the memory kernel as a continuous superposition of exponential kernels,
	\begin{equation}
		h(t) = \frac{1}{n}\int_0^\infty \frac{n(x)}{x}e^{-t/x}dx~, \>\>\> n = \int_0^\infty dxn(x)~
		\label{eq:continuous_decomposition_kernel}
	\end{equation}
	satisfying the normalization $\int_0^\infty h(t)dt=1$
	with the set of continuous time scale $x \in \bm{R}_+ := (0,\infty)$. 
	This decomposition is equivalent to applying the Laplace transform, a standard method even for non-Markovian Langevin equations~\cite{KuboB,Mori1965,Lee1982,Morgado2002} 
	with response functions expanded in terms of a superposition of exponentials~\cite{Bao2006}.
	Here $n(x)$ quantifies the contribution of the 
	$x$-th exponential with memory length $x$ to the branching ratio and 
	$n(x)/n$ is the normalised distribution of time scales present in the memory kernel.
	In this work, we require the existence of its first moment
	\begin{equation}
		\frac{\alpha}{n} :=  \langle \tau \rangle:= \int_0^\infty x \frac{n(x)}{n} dx < \infty~.   \label{rhr2bg2}
	\end{equation}
	This condition (\ref{rhr2bg2}) means that $n(x)$ should decay faster than $1/x^2$ at large $x$'s
	and thus $h(t)$ decays at large times faster than $1/t^2$. 
	In addition, we restrict our analysis to the subcritical case $n<1$.
	
	The starting point of our approach is to express $\hnu(t)$ (\ref{def:Hawkes_general}) 
	as the continuous sum
	\begin{equation}
		\hnu(t) = \nu_0 + \int_0^\infty dx \hz(t,x)~, 
		\label{eq:Markov_exp_pulse}
	\end{equation}
	where each excess intensity $\hz(t,x)$ is the solution of a simple time-derivative equation
	\begin{equation}
		\frac{\partial \hz(t,x)}{\partial t} = - \frac{\hz(t,x)}{x} + \frac{n(x)}{x}\hxi^{\mrP}_{\hnu}(t), \>\>\> \forall x \in \bm{R}_+,
		\label{eq:SPDE_generalHawkes}
	\end{equation}
	and the same state-dependent Poisson noise $\hxi^{\mrP}_{\hnu}(t)$, defined by
	\begin{equation}
		\hxi^{\mrP}_{\hnu}(t) = \sum_{i=1}^{\hN(t)} \delta(t-\htt_i),  \label{hwtrgfq}
	\end{equation}
	acts on the Langevin equation (\ref{eq:SPDE_generalHawkes}) for each excess intensity $\hz(t,x)$ (see Fig.~\ref{fig:Mapping}b).
	The excess intensity $\{\hz(t,x)\}_{x\in \bm{R}_+}$ can be viewed
	as a one-dimensional field variable distributed on the $x$-axis;
	correspondingly, Eq.~\eqref{eq:SPDE_generalHawkes} should be considered as 
	a stochastic partial differential equation (SPDE) describing the classical stochastic dynamics of the field.
	This interpretation has the advantage of allowing us to apply functional methods available for SPDEs~\cite{GardinerB}.
	The introduction of the  $\hz(t,x)$'s is 
	called Markovian embedding, a technique to transform a non-Markovian dynamics onto a Markovian one by adding a sufficient number of variables 
	(see~\cite{Goychuk2009,Kupferman2004,Marchesoni1983} for the cases of non-Markovian Langevin equations). Markovian embedding is 
	related to the trick proposed 
	in \cite{BouchaudTradebook2018} for an efficient estimation of the maximum likelihood of the Hawkes process.
	Each SPDE~\eqref{eq:SPDE_generalHawkes} describes a Markovian relaxation of the field variable $\hz(t,x)$,
	hit by intermittent simultaneous shocks $\hxi^{\mrP}_{\hnu}(t)$ with $x$-dependent sizes $n(x)/x$, 
	whose influence decays exponentially with the characteristic time $\tau$.
	Equation~\eqref{eq:SPDE_generalHawkes} together with (\ref{hwtrgfq}) implies that $\hnu(t)$ given by 
	(\ref{eq:Markov_exp_pulse}) recovers the standard Hawkes definition (\ref{def:Hawkes_general}).
		
	We have thus transformed a non-Markovian point process into a Markovian SPDE, which 
	allows us to derive the corresponding master equation for the probability density functional 
	$P[\{\hz(t,x)=z(x)\}_{x \in \bm{R}_+}] = P_t[z]$ for any field configuration $\{z(x)\}_{x\in\bm{R}_+}$, such that
	$P_t[z]\mathcal{D}z$ is the probability that the system is in the state specified by $\{z(x)\}_{x \in \bm{R}_+}$ 
	at time $t$, with the functional integral volume element $\mathcal{D}z$.
	The corresponding master equation for the probability density functional $P_t[z]$ reads
	\begin{align}
		\label{eq:master_gen_functional}
		&\frac{\partial P_t[z]}{\partial t} = \int dx\frac{\delta }{\delta z} \left(\frac{z}{x}P_t[z] \right) + \\ \notag
		&\left\{\nu_0+\int dx \left(z- \frac{n}{x}\right) \right\}P_t\left[z-\frac{n}{x}\right] - \left\{\nu_0+\int dx ~z  \right\}P_t[z]
	\end{align}
	with the condition $P_t[z]=0$ holding over
	the boundary of the function space $z \in \partial \bm{R}^{\infty}_+:= \{z | z(x = 0 \mbox{ for }x \in (0, \infty) \}$.
	This can be derived by performing an ensemble average in a weak integral sense, namely 
	considering an arbitrary functional $f[\{\hz(t,x)\}_x]$ and averaging it over all possible realisations of $\hz(t,x)$
	weighted by their probability density functional (PDF) (see Ref.~\cite{KiyoDidPRE19} for details).
	The functional description~\eqref{eq:master_gen_functional} is interpreted as a formal continuous limit 
	of a discrete formulation according to the convention~\cite{GardinerB} (see Ref.~\cite{KiyoDidPRE19} for technical details).

	It is convenient to transform (\ref{eq:master_gen_functional}) using 
	the functional Laplace transformation $\Lpath$ of an arbitrary functional $f[z]$ defined by the functional integration (or path integral)
	$\Lpath \big[f[z]; s\big] := \int \mathcal{D}z~ e^{-\int dx s(x)z(x)}f[z]$. Then, 
	the Laplace representation of the probability density functional is $\tl{P}_t[s] := \Lpath \big[P_t[z]; s\big]$
	for an arbitrary nonnegative function $\{s(x)\}_{x\in \bm{R}_+}$. 
	The resulting Laplace transformed master equation (\ref{eq:master_gen_functional}) 
	takes the following simple first-order functional differential equation in the steady state ($\partial{}P_t[z]/\partial{}t=0$):
	\begin{align}
		 \int d\tau \mathcal{H}[s;x]\frac{\delta \Phi[s]}{\delta s(x)} = -\nu_0 \mathcal{K}[s]
		\label{eq:master_gen_functional_cumulant}
	\end{align}
	where  $\Phi[s]:= \log\tl{P}_{\mrss}[s]:= \lim_{t\to\infty}\log\tl{P}_t[s]$ is the steady state cumulant functional,
	$\mathcal{H}[s;x]:= e^{-\int dx' s(x')n(x')/x'} -1 + s(x)/x$, and 
	$\mathcal{K}[s]:= e^{-\int dx' s(x')n(x')/x'} -1$.
	This hyperbolic equation can be solved by the method of characteristics and 
	the corresponding Lagrange-Charpit (LC) equations are the following partial-integro equations,
	\begin{align}
		\frac{\partial s(l;x)}{\partial l} = -\mathcal{H}[s;x], \>\>\>\> 
		\frac{\partial \Phi(l)}{\partial l} = \nu_0\mathcal{K}[s], 
		\label{eq:LagrangeCharpitGeneral}
	\end{align}
	where $l$ is the curvilinear parameter indexing the position along a characteristic curve. 	
	The tail of the distribution of intensities $\hnu$ corresponds to the neighbourhood of $s=0$ in the Laplace transform domain 
	(i.e., $\tl{P}_{\mrss}(s)\sim |s|^{\gamma}$ for $s\to 0$ $\Longleftrightarrow$ $P_{\mrss}(\nu)\sim \nu^{-\gamma-1}$ for $\nu\to \infty$~\cite{KlafterB}).
	We first study the subcritical case $n<1$ and then the critical regime $n=1$ via a
	stability analysis of \eqref{eq:LagrangeCharpitGeneral} for small $s$.
	
	Remarkably, the LC equations can be interpreted as a dynamical system where $l$ plays the role of time.
	This mapping allows us to use the standard stability analysis for bifurcations of dynamical systems, particularly for asymptotic analyses near criticality. 
	Indeed, the stability analysis for $s\to0$ corresponds to the long time limit $l\to\infty$ and  
	the critical condition of the original Hawkes process~\eqref{def:Hawkes_general} corresponds to the transcritical bifurcation condition for the dynamical system described by Eq.~\eqref{eq:LagrangeCharpitGeneral}.
	
	\paragraph{Subcritical case $n<1$.} 
		Linearising the LC equation~\eqref{eq:LagrangeCharpitGeneral} yields
		\begin{subequations}
		\begin{align}
			\frac{\partial s(l;x)}{\partial l} &= -\int dx' H(x,x')s(x'), \\
			\frac{\partial \Phi(l)}{\partial l} &= \nu_0\int dx' K(x')s(x')
		\end{align}
		\end{subequations}
		with
		$x' H(x,x'):= \delta(x-x')-n(x')$ and $K(x') := n(x')/x'$. 
		Introducing the eigenvalues $\lambda \geq \lambda_{\min}$ and eigenfunctions $e(x;\lambda)$ of the operator $H(x,x')$, satisfying the relation
		\begin{equation}
			\int dx' H(x,x')e(x';\lambda) = \lambda e(x;\lambda) ~,
		\end{equation}
		we verify that all eigenvalues are real and 
		the inverse matrix of $H(x,x')$, denoted by $H^{-1}(x,x')$, exists and has a singularity 
		at $n=1$ (see Ref.~\cite{KiyoDidPRE19} for the proof), recovering the critical condition of this Hawkes process.
		
		We now introduce a set of variables to obtain a new representation based on the eigenfunctions,
		\begin{equation}
			s(x) = \sum_{\lambda} e(x;\lambda)X(\lambda)\>\>\>
			\Longleftrightarrow \>\>\>
			X(\lambda) = \int dx e^{-1}(\lambda;x)s(x).
		\end{equation}
		Here the inverse matrix $e^{-1}(\lambda;x)$ is introduced, satisfying $\int dx e^{-1}(\lambda;x)e(x;\lambda') = \delta_{\lambda,\lambda'}$. 
		The existence of the inverse matrix is equivalent to the assumption that the set of all eigenfunctions is complete, 
		and thus $H(x,x')$ can be diagonalized: $\int dx dx' e^{-1}(\lambda;x)H(x,x')e(x';\lambda')  = \lambda\delta_{\lambda,\lambda'}$.
		In this representation, the linearised LC equations read
		\begin{equation}
			\frac{\partial X(l;\lambda)}{\partial l}=-\lambda X(l;\lambda).\label{eq:linearLC_eigenRepresentation}
		\end{equation}
		For subcriticality, all the eigenvalues are positive, indicating that the fixed point $\{X(\lambda)=0\}_{\lambda}$ 
		(i.e., $\{s(x)=0\}_{x}$) is the stable attractor in the functional space. 
		Using straightforward calculations (see Ref.~\cite{KiyoDidPRE19}), we obtain
		\begin{equation}
			\Phi[s] \simeq -\nu_0 \int dx \int dx' K(x)H^{-1}(x,x')s(x'),
		\end{equation}
		from which we find, for small $s$,
		\begin{equation}
			\log \tl{P}_{\mrss}(s) := \log \tl{P}_{\mrss}[s\bm{1}(x)] = \Phi[s\bm{1}(x)] \simeq \frac{-\nu_0}{1-n}s.
			\label{rthr2gqgfq}
		\end{equation}
		where $\bm{1}(\tau)$ is the constant function equal to $1$ for any $\tau$.
		The mean intensity thus converges at long times to $\la \hnu(t)\ra \to \nu_0 / (1-n)$, which is a well-known result \cite{DalayVere03,HelmsSor02}.
					
	\paragraph{Critical case $n=1$.}
		\begin{figure*}
			\centering
			\includegraphics[width=180mm]{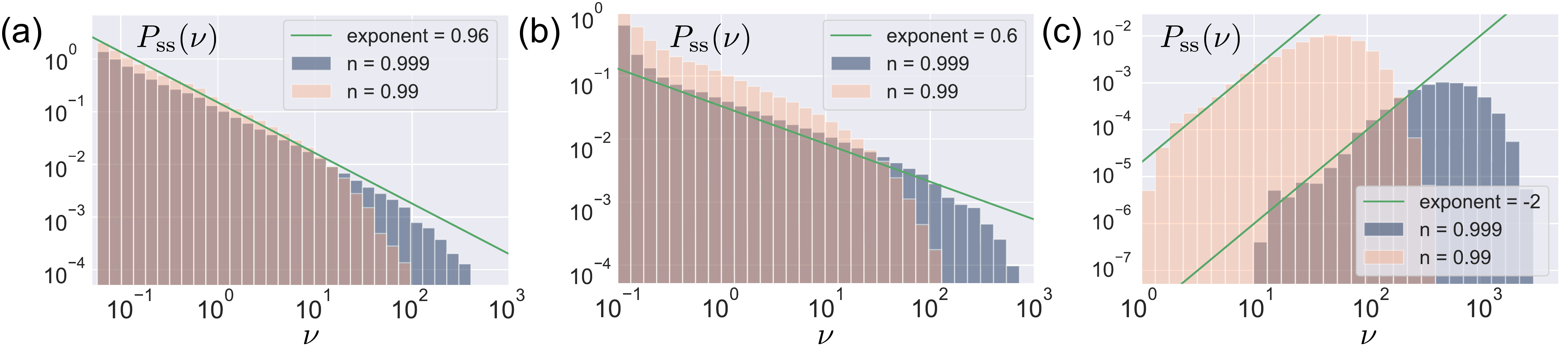}
			\caption{ 
					Numerical steady state PDFs of the Hawkes intensity $\hnu$ for the double exponential case with $(\tau_1,\tau_2)=(1,3)$,
					$(n_1,n_2)=(0.5,0.499)$ or $(n_1,n_2)=(0.5,0.49)$, near the critical point:
					(a) Background intensity $\nu_0=0.01$, leading to the power law exponent $0.96$.
					(b) $\nu_0=0.1$, leading to the power law exponent $0.6$. 
					(c) $\nu_0=0.75$, leading to the negative (i.e. growing PDF) power law exponent $-2.0$.
					Here the sampling time interval and total sampling time are $dt=0.001$ and $T_{\mathrm{tot}}=10000$ from the initial condition $\hz(0)=0$.
					The initial 10\% of the sample was discarded from the statistics for initialization. 
			}	
			\label{fig:Simulation_PDFs_DoubleExpon}
		\end{figure*}				
		At criticality, the smallest eigenvalue vanishes, $\lambda_{\min}=0$, which
		is associated to the zero eigenfunction $e(x;\lambda=0) = x$, as verified by direct substitution: $\int d\tau H(x,x')e(x';\lambda=0)=1-n=0$.
		From the linear LC equation~\eqref{eq:linearLC_eigenRepresentation}, it is clear that the dominant contribution comes from the component $X(\lambda=0)$, associated with the zero eigenfunction $e(x;\lambda=0)$. 
		The explicit representation of $X(\lambda=0)$ is given by $X(\lambda=0) = \int_0^\infty dx n(x)s(x)/\alpha$,
		where $\alpha$ is defined by expression~\eqref{rhr2bg2}.

		We also obtain the LC equations for each component to leading order for $\lambda'\neq 0$, 
		\begin{equation}
			\frac{\partial X(l;0)}{\partial l} \simeq -\frac{X^2(l;0)}{2\alpha}, \>\>\> \frac{\partial X(l;\lambda')}{\partial l} = -\lambda' X(l;\lambda').
		\end{equation}
		This is the normal form of transcritical bifurcations, leading to a log-type singularity in the cumulant for small $s$.
		Indeed, after straightforward calculations, we obtain
		\begin{equation}
			\log \tl{P}_{\mrss}(s) := \log \tl{P}_{\mrss}[s\bm{1}(\tau)]  \simeq \nu_0 s -2\nu_0 \alpha \log |s|,
		\end{equation}
		which by inverse Laplace transform yields
		\begin{equation}
			P_{\mrss}(\nu) \sim \nu^{-1+2 \nu_0 \langle \tau \rangle}~,
			\label{eq:main_finding_power-law_gen}
		\end{equation} 
		using definition (\ref{rhr2bg2}). The exponent $1- 2 \langle \tau \rangle \nu_0$ of the PDF is 
		non-universal and a function of the background intensity $\nu_0$ 
		of the Hawkes intensity and of the average time scale of the memory kernel $\langle \tau \rangle$. 
		As the tail exponent is smaller than $1$, the steady-state PDF $P_{\mrss}(\nu)$ would be not normalizable
		in absence of some cut-off~\footnote{The cutoff tail is typically characterized by an exponential (e.g., see Eq.~\eqref{eq:power-law_single_expon_steady} and Ref.~\cite{BouchaudTradebook2018} for the exponential and power-law memory kernel cases, respectively).}, coming either from finite-time effects or non-exact criticality ($n \to 1^-$).
		This means that this power-law scaling~\eqref{eq:main_finding_power-law_gen} actually corresponds to an intermediate asymptotics of the PDF, 
		according to the classification of Barenblatt~\cite{Barenblatt}, which, for $n$ close to $1$,
		can be observed over many orders 
		of magnitude of the intensity for near-critical systems, as shown in figure \ref{fig:Simulation_PDFs_DoubleExpon}.
		The intermediate power law asymptotic \eqref{eq:main_finding_power-law_gen} is our main novel quantitative result.
		Interested readers are referred to Ref.~\cite{KiyoDidPRE19} for details.

	\paragraph{Example 1.}
		\begin{figure*}
			\centering
			\includegraphics[width=170mm]{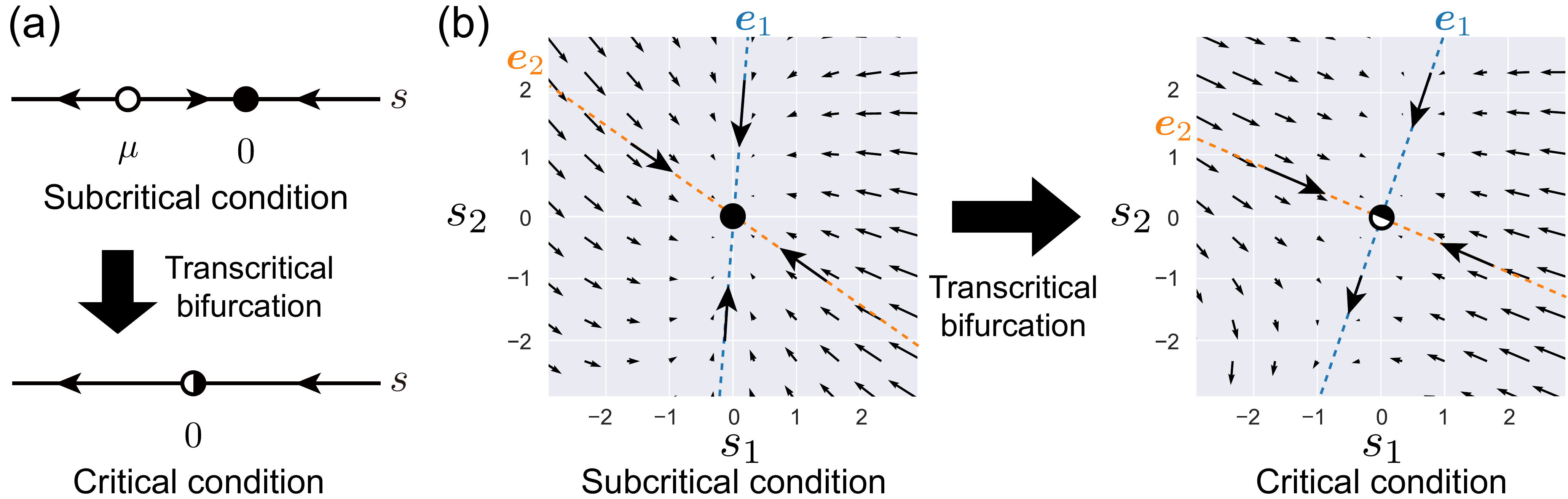}
			\caption{
						Phase space description of the dynamical system following the LC equations. 
						(a)~The one-dimensional velocity field is defined by $V(s):=ds/dl=-\mathcal{H}(s)$. 
						For subcriticality ($n<1$, top), there are two fixed points at $s=0$ (attractor) and $s=\mu<0$ (repeller).
						The repeller merges with the attractor at criticality ($n=1$, bottom), which is a consistent picture of transcritical bifurcations. 
						(b)~The two-dimensional velocity field is defined by $\bm{V}(\bm{s}):=d\bm{s}/dl$ with $\bm{s}:=(s_1,s_2)$.
						By linearisation $\bm{V}(\bm{s})\simeq -\bm{H}\bm{s}$, the eigenvectors $\bm{e}_1$ and $\bm{e}_2$ are introduced by $\bm{H}\bm{e}_i=\lambda_i\bm{e}_i$ with $0\leq\lambda_1<\lambda_2$.  
						For subcriticality ($n<1$, left), the origin $\bm{s}=\bm{0}$ is the stable attractor. 
						At criticality ($n=1$, right), the origin becomes marginal in terms of the linear stability ($\lambda_1=0$): 
						a repeller merges with the attractor along the $\bm{e}_1$ (i.e., a transcritical bifurcation). 
					}
			\label{fig:phaseSpace}
		\end{figure*}

		The above general derivation of (\ref{eq:main_finding_power-law_gen}) is rather involved and one
		can develop more intuition by studying simplest cases where the memory function $h(t)$ is a single exponential or the sum
		of two exponentials. In the former case $h(t)=(1/\tau)e^{-t/\tau}$, all functions become single variables and functional derivatives and integrations
		become standard derivative and integration operators. Then, the general master equation~\eqref{eq:master_gen_functional} reduces to
		\begin{align}
			\frac{\partial P_t}{\partial t} = \frac{1}{\tau}\frac{\partial }{\partial z}zP_t + 
			\left(\nu_0+z-\frac{n}{\tau}\right)P_t\left(z-\frac{n}{\tau}\right)-(\nu_0+z) P_t, 
			\label{eq:master_exp}
		\end{align}
		for the probability density function $P_t:= P_t(z)$	under the boundary condition $P_t(z)|_{z=0}=0$.
		Its Laplace transform of the steady-state PDF $\tl{P}_{\mrss}(s):=\int_{0}^\infty d\nu e^{-s\nu}P_{\mrss}(z)$ reads
		\begin{equation}
			\mathcal{H}(s)\frac{d\Phi(s)}{ds} = -\nu_0 \mathcal{K}(s), \label{eq:maseter_single_expon_steady}
		\end{equation}
		by introducing the cumulant function $\Phi(s):=\log\tl{P}_{\mrss}(s)$, $\mathcal{H}(s):=e^{-ns/\tau}-1+s/\tau$, and $\mathcal{K}(s):=e^{-ns/\tau}-1$.
		It can be directly solved exactly below the critical point $n<1$, leading to
		\begin{equation}
			P_{\mrss}(\nu) \propto \nu^{-1+2n \nu_0 \tau}~ e^{-2 \tau (1-n) \nu}    \label{eq:power-law_single_expon_steady}
		\end{equation}
		for large $\nu$ near criticality $1-n\ll 1$.
		Remarkably, the LC equation $ds/dl=-\mathcal{H}(s)$ reduces to the normal form of transcritical bifurcations (see Fig.~\ref{fig:phaseSpace}a): 
		\begin{equation}
			\frac{ds}{dl'} = \mu s - s^2 + O(s^3)
		\end{equation}
		for small $s$ with $l':=n^2l/(2\tau^2)$ and $\mu:=-2(1-n)/n^2$.

	\paragraph{Example 2.}
		For two exponentials, the memory kernel is given by $h_t=\sum_{k=1}^2 (n_k/(n \tau_k))e^{-t/\tau_k}$, where 
		each coefficient $n_k$ quantifies the contribution of the $k$-th exponential with memory length $\tau_k$
		to the branching ratio $n = n_1+n_2$. 
		In calculations paralleling those for the general and one exponential cases, 
		we can derive the master equation for the two-exponential case and its Laplace representation. 
		Finally, the corresponding LC equations read
		\begin{equation}
			\label{eq:LagrangeCharpit_2expon}
			\frac{ds_i}{dl} = -\mathcal{H}_i(s_1,s_2), \>\>\>\>
			\frac{d\Phi}{dl} = \nu_0 \mathcal{K}(s_1,s_2)
		\end{equation}
		with $i=1,2$,
		$\mathcal{H}_i(s_1,s_2):=e^{-\sum_{k=1}^2n_ks_k/\tau_k}-1+s_i/\tau_i$, and $\mathcal{K}(s_1,s_2):=e^{-\sum_{k=1}^2n_ks_k/\tau_k} - 1$. 
		Following the same approach as for the general case (\ref{eq:continuous_decomposition_kernel}), but now
		dealing with operators that are $2 \times 2$ matrices, we recover (\ref{eq:main_finding_power-law_gen}) with 
		$\langle\tau\rangle=(n_1\tau_1+n_2\tau_2)/n$ (see \cite{KiyoDidPRE19} for details).
		We have numerically confirmed our theoretical prediction for a memory kernel with two exponentials, as shown in
		Fig.~\ref{fig:Simulation_PDFs_DoubleExpon}. 				
		We note that the LC equation~\eqref{eq:LagrangeCharpit_2expon} exhibits the transcritical bifurcation as illustrated in Fig.~\ref{fig:phaseSpace}b.

	The Hawkes process was believed to be unable to reproduce the large fluctuations that are ubiquitously observed in complex systems~\cite{BouchaudTradebook2018}. 
	Our finding demonstrates in fact that the Hawkes process does produce large fluctuations in the form of intermediate asymptotics, thus filling 
		an important gap for applications to real systems. 
	We note that our methodology can be readily generalized to various non-linear Hawkes processes, where wider class of power laws can be discussed~\cite{KanazawaNLHawkes2020}.
	In addition, our main result fills a gap in the study of the Hawkes and other point process, by focusing 
	on the distribution of the number $\nu dt$ of
	events in the limit of infinitely small time windows $[t, t+dt]$. This limit is in contrast to the other previously studied
	limit of infinitely large and finite but very large time windows, as standard results of branching processes (of which the 
	Hawkes model is a special case) give the total number of events generated by a given triggering event  (see Ref.~\cite{SaiHSor2005} for
	a detailed derivation and \cite{SaiSor2006} for the case of large time windows $[t, t+T]$, i.e., in the limit of large $T$'s).
	The corresponding probability density distributions are totally different from (\ref{eq:main_finding_power-law_gen}) 
	which corresponds to the other limit $T \to 0$. 
	There are also deep relationship between our theory and quantum field theories. 
	Indeed, our field master equation can be formally regarded as a Schr\"{o}dinger equation for a non-Hermitian quantum field theory (see 
	derivation in Ref.~\cite{KiyoDidPRE19}), 
	considering the parallel structure between the Fokker-Planck (master) and Schr\"{o}dinger equations~\cite{RiskenB}.

\begin{acknowledgements}
	This work was supported by the Japan Society for the Promotion of Science KAKENHI (Grand No.~16K16016 and No.~20H05526) and Intramural Research Promotion Program in the University of Tsukuba. 
\end{acknowledgements}

%-----------------------------------------------------------------------------------------------------------------------------------------------------------------------------------

%----------------
\end{document}